\documentclass[prl,showpacs]{revtex4-1}
\usepackage{hyperref}
\usepackage{natbib}

\begin{document}

\title{Mass of a Black Hole Firewall}

\date{\today}

\author{M.A. Abramowicz$^{1,2}$}
\email[]{marek.abramowicz@physics.gu.se}
\author{W. Klu\'zniak$^1$} 
\email[]{wlodek@camk.edu.pl}
\author{J.-P. Lasota$^{3,1,4}$}
\email[]{lasota@iap.fr}
\affiliation{$^1$Copernicus Astronomical Center, ul. Bartycka 18, 00-716 Warszawa, Poland}
\affiliation{$^2$Department of Physics, University of Gothenburg, SE-412-96 G{\"o}teborg, Sweden}
\affiliation{$^3$IAP, UMR 7095 CNRS, UPMC Univ Paris 06, 98bis Bd Arago, 75014 Paris, France}
\affiliation{$^4$Astronomical Observatory, Jagiellonian University, ul. Orla 171, 30-244 Krak{\'o}w, Poland}

\begin{abstract}
Quantum entanglement  of Hawking radiation has been supposed to give
rise to a Planck density ``firewall'' near the event horizon of old
black holes.  We show that Planck density firewalls are excluded by
Einstein's equations for black holes of mass exceeding the Planck
mass.  We find an upper limit of $1/(8\pi M)$ to the surface density of a
firewall in a Schwarzschild black hole of mass $M$, translating for
astrophysical black holes into a firewall density smaller than Planck density
by more than 30 orders of magnitude.
A strict upper limit on the firewall density is given by the
Planck density  times the ratio $M_{\rm Pl}/(8\pi M)$.
\end{abstract}
\maketitle

\section{Introduction}
Almheiri  et al.\cite{AMPS}
argue that two standard assumptions made in discussions of quantum
properties of black holes, namely that ``i) Hawking radiation is in a
pure state, ii) the information carried by the radiation is emitted
near the horizon, with low energy effective field theory valid beyond
some distance from the horizon,'' are incompatible with a statement
that iii) the infalling observer encounters nothing unusual at the
horizon \cite{STU}.  Their proposed ``resolution [of the apparent
  contradiction] is that the infalling observer burns up at the
horizon.'' Specifically, they suggest that ``the infalling observer
encounters a Planck density of Planck scale radiation and burns up.''
This phenomenon has been called a ``firewall,''
and firewalls were supposed
to be present both in stellar-mass and supermassive black holes.
While the presence of a firewall is subject to an ongoing controversy
(e.g., \cite{MT}), the internal contradiction of the set of three
assumptions seems to be real and supported by detailed calculations in
\cite{AMPS} and numerous other papers (e.g., \cite{BS} and references
therein). 

With the caveat that the final status of such calculations
is subject to verification in a future theory of quantum gravity, we
would like to explore some consequences of the postulated presence of
a Planck density firewall. 
We will restrict ourselves to the
Schwarzschild black hole, which guarantees spherical symmetry of the
firewall, if present.
As far as we are aware there is no theory of what a firewall is or
what its influence on the (classical) space-time structure would be.
In the present letter we address a specific aspect of this problem
by calculating the gravitational mass of a Planck-density firewall 
in a Schwarzschild black hole.
We take ``Planck density'' to mean a density of $M_{\rm Pl}c^2/l_{\rm Pl}^3$,
with $l_{\rm Pl} =M_{\rm Pl}G/c^2$, and $M_{\rm Pl}=2.18 \cdot10^{-5}\,$g the Planck
mass.  Henceforth, we suppress factors of $G$ and $c$.

\section{Mass of black hole firewalls}

{We begin by checking whether the mass of a firewall may be neglected
in the discussion of the space-time metric.}
The gravitational mass of a spherical shell in the presence of
a Schwarzschild black hole has been examined by a number of authors
(e.g., \cite{FHK}),
 the calculations are a standard application of Einstein's equations and
have never been subject to controversy.
 In the limit of low mass-energy of the shell, i.e., 
neglecting non-linear effects, its gravitational
effect is to increase the mass of the black hole, as perceived by
a distant observer, by $ { M_F=}E_F g_{tt}^{1/2}$, as would be expected from
a simple redshift of energy.
Here, $E_F$ is the mass of the shell in the local frame, and
\begin{equation}
g_{tt}^{1/2}=(1-2M/r)^{1/2}
\end{equation}
is the usual redshift factor associated with a black hole mass $M$
and radial Schwarzschild co-ordinate $r$.
Thus, the gravitational mass of the black hole with the shell placed at $r$ is
\begin{equation}
M_t=M+E_F (1-2M/r)^{1/2},
\label{linear}
\end{equation}
as long as $M\gg |M_t-M|$. 
In the firewall literature everyone seems to be assuming that
{the presence of the firewall would not significantly affect the space-time metric.}
 In fact this is not the case. 
{We begin by showing that a Planck density shell would have a non-negligible  contribution to the gravitational mass of the system.}

It is straightforward to compute the mass-energy of the firewall in the local
rest-frame. We assume a Planck density shell of radius $r=2M$ and
thickness $\lambda$,
equal to the Planck length, $\lambda=l_{\rm Pl}$. Hence, 
\begin{equation}
E_F=16\pi M^2 l_{\rm Pl}{M_{\rm Pl}\over l_{\rm Pl}^3}=16\pi M{M\over M_{\rm Pl}}.
\label{eq:shell_energy}
\end{equation}
The corresponding mass of the firewall, as perceived by a distant observer
would be
\begin{equation}
M_F = 16\pi M{M\over M_{\rm Pl}}(1-2M/r)^{1/2},
\label{eq:linear_mass}
\end{equation}
as long as $M_F\ll M$.
For definiteness we will assume that the firewall is located just
a Planck proper length away from the horizon.
Neglecting a factor of 1/3 which would arise from averaging over the interval
$r=(2M,2M+l_{\rm Pl})$,
the corresponding redshift factor may be evaluated with 
$$
r - 2M = \frac{1}{8M}l_{\rm Pl}^2, 
$$
to yield the final formula
\begin{equation}
M_F = 4 \pi M.
\label{semifinal}\end{equation}
Therefore $M_F > M$, i.e., the firewall would have a mass exceeding the mass
 of the ``bare'' black hole,
 {\ in contradiction with our assumption that $M_F\ll M$}.
A Planck density firewall cannot be treated as a negligible perturbation
to the space-time structure of a black hole.
Therefore, we must consider the ``back reaction'' of the firewall
 on the metric.

We will now examine constraints on the gravitational mass of a black hole and a
shell of energy $E_F$ placed at $r$.
By ref. \cite{FHK} one can express the { gravitational mass} of the shell
 as seen by an observer at infinity as
\begin{equation}
\label{eq:shell_mass}
M_F=M_t-M=E_F\left[\left(1 - \frac{2M}{r}\right)^{1/2} - \frac{E_F}{2r}\right].
\end{equation}
This equation expresses the non-linearity of the Einstein equations \cite{FHK}.
It can be rewritten as
\begin{equation}
\label{eq:quadratic}
\left(\frac{M_F}{E_F}\right)^2
-\left(1 - \frac{2M}{r}\right)^{1/2}\left(\frac{M_F}{E_F}\right)
 + \frac{M_F}{2r}=0,
\end{equation}
with the solution
\begin{equation}
\label{eq:solution}
2\left(\frac{M_F}{E_F}\right) = \left(1 - \frac{2M}{r}\right)^{1/2} + \left(1 - \frac{2M_t}{r}\right)^{1/2}.
\end{equation}
For  $M_t=M$ this reduces to Eq.~(\ref{linear}).
 However, with the shell placed at $r=2M_t$ 
 the equations require $M_t>M$. Indeed, for $r=2M_t$,
 Eq.~(\ref{eq:shell_mass}) reads 
\begin{equation}
\left(\sqrt{M_FM_t^{\ }}-{{E_F}/{2}}\,\right)^2=0,
\label{firemass}
\end{equation}
with the solution
\begin{equation}
\label{eq:mtot}
M_t=\left(M + \sqrt{M^2 + E_F^2}\right)/2.
\end{equation}
Note that $M_t$ does not depend on the sign of $E_F$.
For a fixed value of $E_F$, this would be the final expression for the total
mass of the shell and the black hole.
{In particular, one would have
$M_t-M\gg M$ for $E_F\gg M$. However, the real difficulty with
the firewall is that its energy is proportional to the area of the 
horizon, i.e.,}
 to the square of the black hole mass, {whereas Eq.~(\ref{eq:mtot})
tolerates at most the first power of $M_t$ in the limit of large $E_F$}.
For a Planck density shell at $r=2M_t$
\begin{equation}
\label{eq:shell_energy2}
E_F =16\pi {M_t^2\over M_{\rm Pl}},
\end{equation}
and this makes Eq.~(\ref{eq:mtot}) equivalent to
\begin{equation}
M_t^2 ={M_{\rm Pl}^2\over (8\pi)^2}\left(1-{M\over M_t}\right),
\end{equation}
yielding an upper limit to the total mass of the black hole with
 a Planck density firewall at the horizon
\begin{equation}
\label{limit}
M_t<{M_{\rm Pl}\over 8\pi}.
\end{equation}

We conclude that a firewall of Planck density placed at the horizon of
a Schwarzschild black hole is not a solution to Einstein's equations
for a black hole of mass exceeding the Planck mass.

\section{Discussion}
Our calculations were based on the assumption of a Planck density shell in a static observer's frame.  To complete our discussion of the firewall mass we must examine alternate possibilities to the one just considered. These are

a) the firewall contains negative energy states,

b) the Planck density is measured
in the frame of an observer falling in from infinity,

c) the firewall density is substantially smaller than the Planck density.

\subsection{Negative energy states}
Quantum states can violate the weak energy condition so one may be tempted to take $M_F<0$. Survival of the black hole could then require considerable fine tuning. Indeed, since $|M_F|\gg M$ in Eq.~\ref{eq:linear_mass}, one may worry that a negative value of $E_F$ could lead to an instant disappearance of the black hole. We can venture no opinion on the fate of an observer encoutering a negative energy firewall. Does the observer ``burn up'' or suffer hypothermia?
However, these speculations may be unfounded. Bekenstein \citep{B} argues that negative energy states are associated with positive mass.
If this is the case, our results of Eq.~(\ref{eq:mtot}), and following, remain unchanged in the presence of negative energy states. Such a conclusion would be
consistent with the fact that
$M_t$ in Eq.~(\ref{eq:mtot}) depends on the square of $E_F$.

\subsection{Observer falling in from infinity}
If it is an infalling observer that encounters a Planck density firewall, the energy of the firewall seen by a static observer would be reduced by the appropriate Lorentz factor. However, this would not lead to a decrease of the mass of the firewall observed at infinity, $M_F$, which, as we have seen, is the firewall energy $E_F$ suppressed by an appropriate redshift related to the wall thickness---because of Lorentz contraction, the suppression would be weaker if it is an infalling observer that encounters the firewall, as we now show.

Consider an observer freely falling from rest at infinity. The conserved energy (of the infalling observer) per unit mass is $E=u_t=g_{tt}u^t=1$, implying that the Lorentz factor associated with the motion is $\Gamma=g_{tt}^{-1}$. If the firewall has  Planck density
 ($l_{\rm Pl}^{-2}$) and thickness $\lambda_0$ in the infalling oberver's frame, a static observer at the same location would report a density of 
$(\Gamma l_{\rm Pl})^{-2}$ and thickness $\Gamma \lambda_0$. If this is a thin shell at $r$, the firewall energy seen by the static observer would be
\begin{equation}
\label{alice}
E_F=4\pi r^2{\lambda_0\over \Gamma l_{\rm Pl}^2}.
\end{equation}
Not having at hand a quantum theory of gravity, we must assume that
 $\lambda_0\ge l_{\rm Pl}$. To minimize $E_F$ we take  $\lambda_0= l_{\rm Pl}$.
The relation between incremental thickness $dl_0$ in the infalling observer's frame,  $dl$  in the static observer's frame, and the Schwarzschild radial coordinate increment is
$\Gamma dl_0 = dl = g_{tt}^{-1/2} dr$
so
$$l_{\rm Pl}=\lambda_0=\int dl_0=\int g_{tt}^{1/2} dr= {2\over 3}rx^{3/2},$$
where the last integration is performed from $r $ to $ r(1+x)$.
Taking, as before, $ r=2M$ we have the redshift factor of the shell
$$x^{1/2}=\left({3l_{\rm Pl}\over 4M} \right)^{1/3},$$
and $\Gamma=x^{-1}$. Eq.~(\ref{linear}) now gives
\begin{equation}
\label{alice2}
M_F=16\pi M{M\over M_{\rm Pl}}x^{3/2} = 12 \pi M,
\end{equation}
a result essentially unchanged from that of Eq.~(\ref{semifinal}).

\subsection{Lower density firewall}
It is true that by sufficiently decreasing the density of the firewall one may avoid the difficulties discussed above. However, on dimensional grounds it is expected that {as a quantum gravity effect} the firewall has a Planck density, so while a lower density firewall could perhaps be made to satisfy Einstein's equations, explaining the magnitude of the density would pose a separate problem for the theory. Further, as we will now show, the firewall density cannot be universal, it has to depend on the black hole mass.

Let us allow the firewall density to be reduced
by a factor of $\beta\Gamma$ with respect to the Planck density,
 where $\Gamma=1$ if the density is chosen in the static observer's frame, and 
\begin{equation}
\label{gamma}
\Gamma=\left({4M\over 3M_{\rm Pl}}\right)^{2/3}
\end{equation}
if the density is chosen in the freely-falling observer's frame.
Repeating the calculation of firewall mass-energy and gravitational mass
for $r=2M_t$ we get
\begin{equation}
\label{eq:shell_energy3}
E_F=16\pi {M_t^2\over M_{\rm Pl}}\beta,
\end{equation}
and
\begin{equation}
M_t^2 ={M_{\rm Pl}^2\over (8\pi\beta)^2}\left(1-{M\over M_t}\right).
\end{equation}
As the firewall is supposed to arise as a result of (dis)entanglement
of Hawking radiation, we assume that the gravitational mass of the black hole
with the firewall should not be appreciably larger than that of the ``bare'' black hole. Accordingly, we take
$M_t$ to differ from $M$ by a quantity on the order of Planck length
\begin{equation}
\label{assume}
M_t -M =\alpha M_{\rm Pl},
\end{equation}
with $\alpha={\cal O}(1)$. Finally, we obtain
\begin{equation}
\beta
 ={\alpha^{1/2}\over 8\pi }\left({M_{\rm Pl}\over M}\right)^{3/2}
 \sim 0.1\left({M_{\rm Pl}\over M}\right)^{3/2}\ll 1.
\label{beta}
\end{equation}
{Eq.~(\ref{firemass}) directly implies
 $E_F=2\sqrt{(M_t-M)M_t}$.
Note that this translates into a limit on the surface density of the firewall of
$E_F/(16\pi M_t^2)<1/(8\pi M_t)$. This is to be compared with the Planck
surface density $1/M_{\rm Pl}$. Since the thickness $\lambda$
of the shell cannot be lower
than the Planck length, we arrive at a strict upper limit on the firewall
density}
\begin{equation}
{E_F\over 16\pi\lambda M_t^2}
< {1\over 8\pi}\left({M_{\rm Pl}\over M_t}\right) M_{\rm Pl}^{-2},
\end{equation}
{i.e., the density has to be diminished with respect to the Planck density
by at least the factor $M_{\rm Pl}/ M_t$.
The additional 1/2 power of this factor in Eq.~(\ref{beta}) follows
from the additional requirement that the firewall
contribute  to the gravitational mass
no more than a quantity on the order of a Planck mass.}

We end this discussion with a caveat.
Eq. (12) is based on the assumption that the firewall
is placed just outside the horizon,
as considered by some authors, e.g., \cite{Hawk}.
After this work was completed it has been brought to our attention
that ``it is generally believed that if the firewall is
real it is restricted to the interior.''
Inside the horizon ($r<2M$) no static structures can exist, and all
geodesics end at the Schwarzschild singularity.  Quite apart from
the problem of the firewall energy  greatly exceeding
the mass of the black hole, c.f. Eq. (11),
which necessitates a careful treatment
of the influence of the firewall on the space-time metric,
placing a steady firewall inside the horizon would give rise to
the additional difficulty of the black hole mass varying in time
as the firewall photons accrete onto the central singularity
while the firewall is continually being recreated. A computation of the
resulting rate of change of the black hole mass is outside the scope of
this {\sl Letter}.
These difficulties could possibly be cured if the firewall existed deep
inside the black hole at, say, $r<<\sqrt{M l_{\rm Pl}/ 16\pi}$.

\section{Conclusions}

We have shown that a Planck density firewall placed at the Schwarzschild horizon does not satisfy Einstein's equations for a black hole mass exceeding the Planck mass. Any shell located at the horizon of an astrophysical black hole must necessarily have a density many orders of magnitude lower than the Planck density.

We find that the density of the firewall is at most
$\beta\Gamma M_{\rm Pl}^{-2}$, with $\beta$ given
by Eq.~(\ref{beta}), and $\Gamma$ by Eq.~(\ref{gamma}),
its mass-energy being proportional to the geometrical mean of the Planck mass
 and the gravitational mass of the system $E_F=2\sqrt{\alpha M_{\rm Pl} M_t}$.
For a  $10M_\odot$ black hole the density works out to be $\beta\sim10^{-59}$
 Planck density if it is discussed in the static observer's
 frame,
or $ \beta\Gamma\sim(M_{\rm Pl} /M)^{5/6}\sim 10^{-34}$ Planck density
 if the density is discussed in the infalling observer's frame. 
These densities are considerably smaller than the ones assumed in
 \cite{AMPS,Br}.

\bigskip
\begin{acknowledgments}
We are grateful to Drs. Don Marolf and Iwona Kotko for helpful comments.
We thank the referees for critical remarks that helped us to improve the
presentation of the results.
This work was supported in part by Polish NCN grants 2013/08/A/ST9/00795,
and 2011/01/B/ST9/05439 and by the National Science Foundation
under Grant No. NSF PHY11-25915.
\end{acknowledgments}

\end{document}